# Physics of quantum mechanics - why are there phonons?

R. A. Street

Palo Alto Research Center, Palo Alto, CA 94304

The underlying physics of quantum mechanics has been discussed for decades without an agreed resolution to many questions. The measurement problem, wave function collapse and entangled states are mired in complexity and the difficulty of even agreeing on a definition of a measurement. This paper explores a completely different aspect of quantum mechanics – the physical mechanism for phonon quantization – to gain insights into quantum mechanics that may help address the broader questions.

Ever since quantum mechanics became established as a theory that correctly predicts the result of a measurement, there has been a continuing discussion about the underlying physics and the meaning of the theory. Most of the discussion is centered around the measurement problem, decoherence, entangled states and the nature of the wave function.[1-6] The measurement problem concerns how a quantum state interacts with the macroscopic external world to give the observed probabilistic result of a measurement. The problem focuses on reconciling the unitary evolution of the quantum state and the non-unitary result of a measurement. This dichotomy has been under discussion for nearly a century without a clear resolution and is complicated by the difficulty in even finding a precise and agreed definition of a measurement. The wave function is similarly complicated by a lack of clarity whether it reflects a real physical entity or is a mathematical abstraction.[7,8] These questions of ontology, epistemology and broader philosophical issues have proved very hard to unravel, in part because of the lack of knowledge about the underlying physics of quantum mechanics.[9-13] A core issue is one of ontology and the nature of reality. The realist point of view is that there must be a physical entity that is the source of the quantum properties.

Given the difficulties described above, it seems helpful to try to make progress with a potentially simpler question. To that end, this paper considers an aspect of quantum mechanics that is rarely discussed in the context of the foundations of quantum mechanics, namely phonons, and to ask why they are quantized, obeying the $E=(n+1/2)\hbar\omega$ energy-frequency (E-ω) relation. Phonons are relatively simple constructs, being mechanical oscillations of a group of atoms. The point of taking this approach is that the measurement problem and the nature of the wave function are peripheral to the description of phonons, but perhaps some insight can be gained that will help with those questions. The mathematical description of phonons as the quantization of a harmonic oscillator is well known but does not explain the quantization. Phonons have a specific property that is pertinent to their physical origin; a phonon mode involves multiple particles, so that one cannot attribute their quantization to the property of any one specific particle, as discussed further below. Phonons are often macroscopic oscillations, examples being acoustic phonons in a crystal and the nano-mechanical oscillation of a $C_{60}$ molecule.[14] The relation of phonons to macroscopic material properties is also clear and direct – the low specific heat of materials at low temperature is the result of phonon quantization. The quantum mechanics of macroscopic objects is of general interest.[15]

The question being addressed is the physical origin of phonon quantization – i.e. what is the specific physical entity whose properties are the origin of the quantization? It is worth starting by reviewing the history. As is well known, Einstein proposed that lattice vibrations are quantized in a 1907 paper and explicitly used the result to explain why the specific heat of materials fell below the Dulong-Petit classical value at low temperature.[16] However, Einstein did not propose a physical mechanism for phonon quantization, but instead gave an argument that the existence of quantized photons implied that phonons were also quantized.

In 1900, Plank showed that the theory of black body radiation required the $E=\hbar\omega$ relation to account for the spectrum.[17] Initially Planck did not consider that the quantum relation reflected new physics and he spent several years trying to find a classical explanation.[18] It was not immediately obvious that new physics must be involved because the calculation of the black body spectrum was not settled, and its theoretical description was being actively debated. Part of Einstein's 1907 paper settled this debate and gave the calculation of the spectrum in the way that it is done now, comprising the density of radiation modes in the cavity, equipartition for modes of the same energy and a thermal distribution of the mode energy. The 1907 paper showed that a Boltzmann energy distribution gives the correct spectrum at low frequency but diverges at high frequency. Einstein identified the classical Boltzmann distribution as the problem and that the correct theory required a distribution function reflecting energy quantization. There was no question that this could be explained by classical physics.

At this point the discussion given by Einstein in the 1907 paper gets more interesting. He could have asserted that photons are intrinsically quantized to explain the quantized energy distribution. (Here, intrinsically quantized means that the quantum effect is a property of the photon independent of the excitation source, rather than a property of the mechanism by which it is excited.) However, while the intrinsic quantization mechanism gives the black body spectrum, it would not have given him an argument for phonons. Instead he asserted that the general oscillatory electronic excitations of the cavity wall material must be quantized, so that only photons with the quantum energy relation can be excited. He then argued that quantized electronic excitations of a material must imply quantized ionic oscillations and therefore also lattice vibrations. He concluded that that phonons must have the same $E=\hbar\omega$ quantum relation as do photons. Einstein's argument is based on the proposition that electronic oscillations of bulk matter have properties that provide the physical origin of quantization. The 1907 paper was not very specific about the mechanism. It states "…the mechanism of energy transfer is such that the energy of elementary structures can only assume the values $(n\hbar\omega)$…" with a footnote "It is obvious that this assumption also has to be extended to bodies capable of oscillation that consist of any number of elementary structures".

A century later we know that photons are intrinsically quantized,[19] so that Einstein's mechanism to link photons with phonons is not necessarily correct. However, phonon quantization is a fact and the mechanism must originate in the oscillating matter. One possible explanation is that the quantum properties of individual particles combine to quantize the phonon, but this is readily refuted. To make the reasoning specific, consider a single benzene molecule $C_6H_6$, made up of 40 electrons, 40 protons and about 36 neutrons (without considering quarks) and for which the phonon frequencies are well known. Benzene has 20 distinct vibrational frequencies and hence 20 different energy quanta.[20] For the atoms of the benzene molecule to be the source of the quantization, then some physical property of those 116 particles must somehow generate the energy quanta. The oscillation frequency results from the mass $m$ and elastic force constant $k$, $\omega=(k/m)^{1/2}$. The elastic energy of the phonon is held by the stretched bonds of the valence electrons – different combinations for each phonon energy – while the kinetic energy is all in the mass of the nuclei, and the energy quantum must somehow transfer back and forth since the energy alternates between the elastic and the kinetic components. The fully symmetric phonon modes of benzene have vibrational energy that is shared equally by the 6 CH units. By symmetry, each unit must provide the energy for 1/6 of the quantum. It is not possible from any reasonable understanding of elementary particles and atoms that they could combine their individual quantum properties to explain the phonon quantum. There has to be a different explanation.

The phonon, photon and other oscillations are described by the quantum theory of the harmonic oscillator. The theory treats the frequency as an arbitrary parameter, which just goes to confirm that the physical mechanism of quantization does not lie in the specific mechanisms that determine the oscillation frequency. Instead, the theory shows that the quantum physics lies in the creation and annihilation operators. However, while the mathematics of the harmonic oscillator is clear, the physics that these operators describe is opaque. Neither the momentum operator or the ladder operator version of the theory point to an obvious physical mechanism. For example, the momentum operator *ihd/dx*, describes how the physics leads to the mathematical equations, but does not reveal the physical mechanism.

Faced with this impasse, a helpful feature of the phonon is one cannot readily appeal to some higher level of abstraction, either mathematical, physical or cognitive as is often done to try and understand quantum physics. There is no obvious relevance in considering the phonon in the context of quantum field theory, string theory or any other higher abstraction. The many-worlds interpretation or the role of human consciousness have no apparent relation to phonons, because they are such simple physical structures. They are not sub-atomic or relativistic – they are just a few atoms oscillating slowly, visible and evidently real; benzene molecules are observable in a scanning tunneling microscope.[21]

The central distinguishing feature of the phonon that sets it apart from the quantization of individual particles is that it is a collective property of a cluster of vibrating atoms, a cluster that can contain almost any number of atoms, as noted in the 1907 paper. The only relevant collective physical quantities in such a mechanical vibration are frequency and energy; there is nothing else to consider. The above discussion shows that the frequency is not where the quantum physics originates. Both the mathematical analysis and the physics shows that the vibration frequency is a result of understood atomic forces. The energy is quantized, not the frequency. The conclusion is that the quantum properties must arise from the physical properties of energy. Energy cannot be what we usually think it is, but instead it has physical properties – the quantum properties. $E=\hbar\omega$ describes a property of energy for any oscillatory phenomenon.

The detailed discussion of benzene is intended to make clear that the phonon quantization cannot be attributed directly to the individual particle properties. Once it is accepted that the physical mechanism must instead be a collective property, then energy is the inescapable origin, at least in any realistic ontology – it is the only possibility. It is perfectly reasonable that energy should be a physical entity with a set of physical properties even though this is not the conventional view, and there is no contradiction with any evidence. Energy is an observable; a measureable quantity that reflects the physical properties of a system, either a quantum system or a macroscopic classical system. Its role as an observable does not prevent energy also having physical properties that contribute to the properties of a quantum system. A comparable example is electric charge; the quantity of electric charge is an observable, but charge also has physical properties that determine the behavior of any system that contains charge. In the macroscopic world, the specific properties of energy that lead to quantum effects are too small to be significant and only the numerical value is important. Similarly, in macroscopic situations electric charge is a numerical quantity, amps or coulombs, but at a microscopic level it is "quantized" as the charge on an electron or proton.

Energy provides the direct link between the photon and the phonon that Einstein made indirectly – the common feature is the energy of the oscillation and the $E=\hbar\omega$ constraint placed on it by the properties of energy. Quantum physics concerns the properties of energy; the Schrödinger and Dirac equations, and the Lagrangian of field theory are each descriptions of energy.

Accepting that energy has its own physical properties, it follows that all quantum states comprise two physical components; one is the source of the energy (radiation, lattice vibrations, particles) and the other is the energy state. Such a two-component quantum state has implications for the description and understanding of quantum physics. For example, Bohr introduced the Principle of Complementarity as a way of framing the underlying physics. This is a statement that particles have pairs of complementary properties which cannot both be observed or measured simultaneously. Bohr's formulation of the Principle describes the results of the quantum equations that were developed from experimental results in the context of the Copenhagen interpretation. It is a descriptive statement rather than a physical mechanism. A two-component quantum state provides the underlying physical mechanism; the two components that comprise the quantum state are complementary and in the case of a particle, an experiment either measures the energy state (or a related quantity such as momentum) or measures the particle state, but not both.

The two-component quantum state also explains the quantization of electronic states, one of the many aspects of quantum mechanics that was never previously explainable in physical terms. Although an electron near an atom nucleus can in principle have a continuum of energies depending on the separation distance, only a few quantized energies are allowed. The Schrödinger equation describes the allowed states but does not explain why they occur in physical terms. The explanation provided by the two-component quantum state is that the physical properties of the energy state constrains the Coulomb energy of the electron to a limited number of allowed states consistent with both components of the quantum state. This is a familiar situation in physics. For example, the equations of heat diffusion, hydrodynamics and other processes allow for an infinite number of possible solutions. Unique solutions are obtained only when an additional constraint is added, which is usually a boundary condition. Quantization is not a mysterious discretization of nature but instead a normal result of a constrained system. A recent paper explores in further detail how the properties of energy can account for a variety of quantum properties.[22]

A continuing aspect of the discussion surrounding the fundamentals of quantum properties is the question of ontology versus epistemology – a discussion that has been going on for at least 50 years.[13] The ontology view is that the wave function is real and the Schrödinger equation should fully specify the quantum state.[5,12,23] The problem with this view is that the non-unitary wave function collapse is not described by the unitary Schrödinger equations, apparently contradicting the ontological assumption of fully specifying the quantum state. The epistemological view is that the wave function is not real but instead is an abstraction that reflects the fundamentally limited information that can be known about the quantum state.[7,8] Wave function collapse is no longer a problem because the collapsed wave function reflects the new information that came about from a measurement. However, the epistemological view brings fully into question the nature of reality and leads to troubling alternatives that few physicists are willing to embrace.[5,11,12,13]

The two-component quantum state does not immediately resolve these questions but certainly changes the basis of the discussion. For example, the wave function does not fully describe the quantum state, as assumed by the ontological approach, but instead describes only the energy of the quantum state. The particle is a separate component of the quantum state, linked in some way to the energy component. The energy state wave function constrains the particle to a region of space but does not define its position. Absent the assumption that the wave function fully defines the quantum state, the primary argument against the ontological view is removed. No further arguments are offered here for how the two-component quantum state changes the understanding of wave function collapse, the measurement problem and other aspects of the quantum properties, which await more careful consideration. The key point is that the two-component quantum state must changes the basis of the discussion.

In summary, it is argued that phonon quantization must be the result of the physical properties of energy – no other explanation seems possible. If energy is a real physical entity then quantum states must be understood as being made up of two components with their own specific physical properties. These two components acting together provide a rather straightforward explanation of quantization of electronic states, the Principle of Complementarity and other aspects of quantum mechanics. The two component approach also changes the basis for discussion of the measurement problem and the general questions of ontology and epistemology.